\documentclass[usenatbib]{mnras}
\usepackage[dvips]{graphicx}
\usepackage[english]{babel}
\usepackage{amsmath}
\usepackage{amssymb}
\usepackage{newtxtext,newtxmath}

\newcommand{\Msun}{{\rm M}_{\odot}}

\newcommand{\etaM}{\eta_{\rm M}}
\newcommand{\etaE}{\eta_{\rm E}}
\newcommand{\Mdot}{\dot{M}}
\newcommand{\Mdotw}{\Mdot_{\rm wind}}
\newcommand{\Edot}{\dot{E}}
\newcommand{\Edotw}{\Edot_{\rm wind}}
\newcommand{\Sigmagas}{\Sigma_{\rm gas}}
\newcommand{\Msunpc}{\Msun~{\rm pc}^{-2}}
\newcommand{\kms}{km~s$^{-1}$} 
\newcommand{\vcirc}{v_{\rm circ} }
\newcommand{\rcool}{r_{\rm cool}}
\newcommand{\fcl}{f_{\rm cl}}

\newcommand{\be}{\begin{equation}}
\newcommand{\ee}{\end{equation}}
\newcommand{\bea}{\begin{align}}
\newcommand{\eea}{\end{align}}
\newcommand{\beg}{\begin{gather}}
\newcommand{\eeg}{\end{gather}}

\defcitealias{CC85}{CC85}
\defcitealias{Martizzi+16}{M16}

\title[How Supernovae Launch Galactic Winds]{How Supernovae Launch Galactic Winds}
\author[Fielding, Quataert, Martizzi, \&  Faucher-Gigu\`ere]{
Drummond~Fielding$^{1}$\thanks{E-mail: dfielding@berkeley.edu}, 
Eliot~Quataert$^{1}$, 
Davide~Martizzi$^{1}$, 
\& Claude-Andr\'e Faucher-Gigu\`ere$^{2}$\\
 $^1$Astronomy Department and Theoretical Astrophysics Center, University of California Berkeley, Berkeley, CA 94720, USA\\
 $^2$Department of Physics and Astronomy and CIERA, Northwestern University, 2145 Sheridan Road, Evanston, IL 60208, USA}

\begin{document}

\date{Accepted not yet. Received \today; in original form \today}

\pagerange{\pageref{firstpage}--\pageref{lastpage}} \pubyear{2017}

\maketitle

\label{firstpage}

\begin{abstract}
We use idealized three-dimensional hydrodynamic simulations of global galactic discs to study the launching of galactic winds by supernovae (SNe).  The simulations resolve the cooling radii of the majority of supernova remnants (SNRs) and thus self-consistently capture how SNe drive galactic winds. We find that SNe launch highly supersonic winds with properties that agree reasonably well with expectations from analytic models. The energy loading ($\etaE =  \Edotw / \Edot_{\rm SN}$) of the winds in our simulations are well converged with spatial resolution while the wind mass loading ($\etaM = \Mdotw/\Mdot_\star$) decreases with resolution at the resolutions we achieve. We present a simple analytic model based on the concept that SNRs with cooling radii greater than the local scale height breakout of the disc and power the wind. This model successfully explains the dependence (or lack thereof) of $\etaE$ (and by extension $\etaM$) on the gas surface density, star formation efficiency, disc radius, and the clustering of SNe. The winds in the majority of our simulations are weaker than expected in reality, likely due to the fact that we seed SNe preferentially at density peaks. Clustering SNe in time and space substantially increases the wind power. 
\end{abstract}


\section{Introduction}

Galactic winds help limit the efficiency with which galaxies turn gas into stars by expelling material from the ISM and by halting gas inflow into galaxies \citep[e.g.,][]{DekelSilk,SpringelHernquist03,CAFG+11}. They are also responsible for enriching and heating the CGM \citep[e.g.,][]{Aguirre+01,OppenheimerDave2006,Hummels+13,Fielding+17}. As a result galactic winds are at the heart of many of the cornerstone relationships of modern astronomy such as the stellar mass function, stellar mass to halo mass relation, and the mass-metallicity relation. Many processes are capable of launching galactic winds, but in star forming galaxies energy deposition by SNe is often thought to be a key driver. 

Constraints on the nature of star formation powered galactic winds come from numerous sources. First, extensive observations have directly measured the energy and mass loading of galactic winds in different environments \citep[e.g.,][]{Heckman+90,Veilleux+05,Chisholm+17}. Second, analytic considerations predict the density, temperature, and velocity profiles of a galactic wind for a given energy and mass loading (e.g., \citealt{CC85} hereafter \citetalias{CC85}, \citealt{Thompson+16}). Third, cosmological simulations demonstrate that winds with a particular range of efficiencies are required to reproduce many observations \citep[e.g.,][]{FinlatorDave08,SomervilleDave15,Muratov+15}. Given all we know about galactic winds there is nonetheless a surprising degree of disagreement on if/how SNe are capable of launching winds that meet all the necessary constraints. 

Numerical simulations of isolated galaxies inform how SNe drive winds---commonly using local stratified box simulations \citep[e.g.,][]{JoungMacLow06,Creasey+13,Girichidis+16, Li+16, Kim+16}. These stratified box simulations generally predict winds that are subsonic, which may be a result of the geometry, namely the lack of a well-defined escape speed and free-space for the wind to expand into (\citealt{Martizzi+16} hereafter \citetalias{Martizzi+16}). 
To more faithfully address how SNe launch galactic winds we designed a new suite of simulations that adopts a global geometry, capturing an entire gaseous galactic disc while resolving most SNRs. Because we neglect self-gravity, molecular line cooling, and other important physics these are not the final word on the true energy and mass loading of SNe-driven winds. But they do significantly sharpen our understanding of the origin and properties of such winds.  

\section{Method} \label{Method}
We ran a series of simulations designed to study the launching of galactic winds by SNe in a global geometry using the Eulerian hydrodynamics code \textsc{athena} \citep{Stone+08}. We evolve a gaseous galactic disc that is stratified by an external potential---representing the gravitational field from baryons and dark matter---in which intermittent, discrete SNe go off at a given rate. The setup of the numerical experiment is simple and provides a useful counterpoint to analogous experiments that differ essentially only in their use of local Cartesian simulation domains; we compare primarily to our earlier work \citepalias{Martizzi+16}. The simulations evolve an ideal fluid in three dimensions with cooling and without self-gravity. The gas has solar metallicity everywhere and cooling proceeds assuming collisional ionization equilibrium. Cooling below $10^4$ K and photoelectric heating are not included.
The gas is initialized in a rotating, $10^4$ K disc that is in radial centripetal balance and vertical hydrostatic equilibrium with the background gravitational potential that is given by a Hernquist profile, $\Phi(r) = - {G M_\circ}/({r+a})$,
where $r$ is the spherical radius ($R$ represents the cylindrical radius) \citep{Hernquist}. The disc initially has an exponentially declining surface density profile characterized by a central surface density $\Sigmagas$ and a scale length $R_d$ that is set to match the characteristic radius of the gravitational potential $a$. 

The simulation parameters are listed in Table \ref{simulations}. In all simulations the parameters are chosen so that $\vcirc(r=a=R_d) = \sqrt{G M(r) / r} =100$ \kms. For our fiducial simulations we adopt the relatively small disc scale length of $R_d = 300$ pc, although we study the differences resulting from using a ${\sim}3\times$ smaller and larger disc. Our disc sizes and $\vcirc$ are similar to those in M82 and NGC 253, but our $\Sigmagas$ are not quite as high.

The major drawback of running global rather than local simulations is the computational expense of resolving the SNR injection when we are primarily interested in the large-scale structure and wind dynamics. This partially motivates the smaller disc scale-lengths considered here, so that SNR evolution can be resolved. To further this goal, we use five levels of static-mesh-refinement. We space our cubical $256^3$ grids logarithmically, enabling us to use a domain side length of $L_{\rm box} \gtrsim 40 R_d$ and have high resolution (3 pc) in the disc mid-plane where most of the mass resides (the highest resolution region is $L_{\rm HR} \gtrsim 2 R_d$ on a side).

\begin{table}
\vspace*{-0.25cm}   
\centering
\begin{tabular}{l|l|l|l|l}
$M_\circ$$[\Msun]$    & $R_d$[pc]& $\Sigmagas$$[\Msunpc]$ & $\fcl$ & $f_\star/100$  \\ \hline
$1.0\times10^{9}$ & 100        & 10, 30, 50                    & 1 & 0.3, 1, 3 \\
$2.8\times10^{9}$ & {\bf 300} & {\bf 10, 30, 100, 300}              & {\bf 1}, 3, 10, 30 & 0.3, {\bf1}, 3 \\
$9.2\times10^{9}$ & 1000      & 10, 30                                    & 1                 & 1       
\end{tabular}
\caption{Full range of simulation parameters. Not all combinations are discussed in the text. The fiducial models are in bold. $M_\circ$ is the mass of the external gravitational potential. $R_d$ is the characteristic radius of the gravitational potential and the scale length of the exponential gas disc. $\Sigmagas$ is central gas surface density. The parameter $\fcl$ is the clustering factor of SNe: the total energy per SNR is equal to $\fcl \times 10^{51}$ ergs while the SN rate $\propto \fcl^{-1}$. The efficiency of star formation is encapsulated in $f_\star = t_\star/t_{\rm dyn}$, such that a lower $f_\star$ yields more star formation. }
\label{simulations}
\vspace*{-.20cm}   
\end{table}

\begin{figure}
\vspace*{-0.25cm}   
\includegraphics[width=0.525\textwidth]{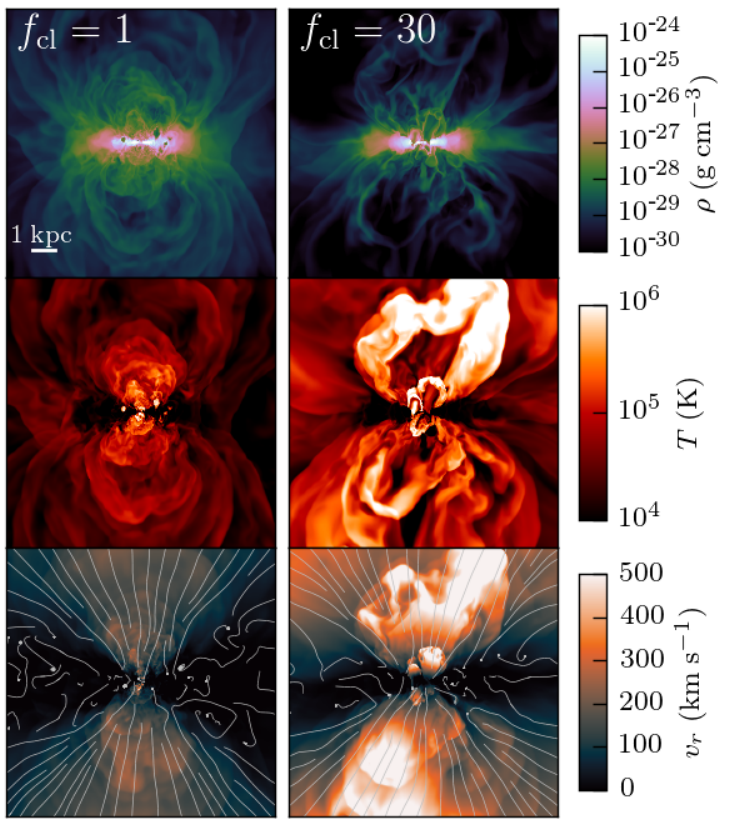}
\vspace*{-0.5cm}   
\caption{Slices along the disc rotation axis showing the density, temperature, spherical radial velocity for the $R_d = 300$ pc and $\Sigmagas = 10~\Msunpc$ simulations with $f_\star = 100$ and $\fcl = 1$ (left column), and $\fcl=30$ (right column). Each image spans the entire computational domain and is 12.3 kpc across. In several places a SNR can be seen breaking out of the disc and powering the wind. The streamlines in the velocity plot trace the flow direction and highlight the nearly straight radial outflow in the central biconical region, turbulence in the disc, and shearing and fountain flow between the two. Clustering the SNe ($\fcl = 30$; see \S\ref{Method}) significantly increases the wind velocity, temperature, and mass and energy outflow rates. (Movies are available at \url{http://w.astro.berkeley.edu/\~dfielding/\#SNeDrivenWinds})}\label{fig:image}
\vspace*{-.05cm}   
\end{figure}

We seed and inject SNRs in our simulations as in the ``SC'' model used by \citetalias{Martizzi+16}, so we refer the reader there for more details. In short, the probability of a SN being set off is proportional to the local gas density and inversely proportional to the local star formation timescale, which is chosen to be proportional to the dynamical time, so that
\be
P({\rm SN\,in\,cell}) = \frac{M_{\rm cell}}{100 \Msun} \frac{dt_{\rm hydro}}{t_\star} \propto \frac{n_{\rm cell}}{t_\star} = \frac{n_{\rm cell}}{f_\star t_{\rm dyn}}.
\ee
This assumes that for every 100 $\Msun$ of stars that form there is one SN. Our fiducial choice of $f_\star$ is 100, which corresponds to a 1\% star formation efficiency and results in star formation rate surface densities $\dot{\Sigma}_\star$ that are similar to observations.

SNRs are injected using the subgrid model developed by \cite{Martizzi+15}, which accounts for subgrid cooling and injects both kinetic and thermal energy at a value calibrated to high resolution single SNR simulations. Additionally, 3 $\Msun$ of ejecta is added to the SNR per SN, so the ISM mass loading ($\Mdot_{\rm ej}/\Mdot_\star$) is 0.03. One of the primary aims of this study is to determine wind mass and energy loss rates when the SNRs' cooling radii ($\rcool$) are explicitly resolved, so we chose parameters to ensure that this occurs for our higher resolution simulations. In this limit our SNe injection model corresponds to $2.9\times10^{50}$ ergs of kinetic energy and $7.1\times10^{50}$ ergs of thermal energy. One new feature we added to the injection scheme relative to \citetalias{Martizzi+16} is a somewhat crude model for the clustering of SNe in space and time (future work will expand this feature). We allow the injected energy per SN to be scaled up by an integer clustering factor $\fcl$, which represents multiple SNe going off simultaneously (the SN rate is correspondingly reduced by $\fcl$, so that the total injected energy by SNe is unchanged). The cooling radius and other radii in the \cite{Martizzi+15} fits for subgrid injection are scaled up by $\fcl^{2/7}$ in accordance with analytic expectations \citep{Cioffi+88}.

\section{Results}\label{Results}

We begin our presentation of the simulation results with a qualitative description to ground the readers' intuitions. In Fig. \ref{fig:image} we show density, temperature, and spherical radial velocity images from simulations with $R_d = 300$ pc, $\Sigmagas = 10~\Msunpc$, $f_\star = 100$ and both $\fcl = 1$ and $\fcl=30$, after 300 Myr of evolution (${\sim}16~t_{\rm orb}$). These images are slices through the computational domain along the rotation axis of the disc. Clearly shown are the strong biconical outflows driven by the SN. Along the rotation axis of the disc densities are low, temperatures are high, and velocities reach upwards of 300 \kms ($\sim 500$ \kms in the clustered model). In the midplane of the disc densities remain high, the temperature remains at roughly $10^4$ K, and the gas is turbulent with a mass-weighted velocity of ${\sim}10$s \kms. These images show several supernova remnants breaking out of the disc. These breakout events carve out a region of the disc and dump thermal energy into the low density wind region, thereby powering the wind. The discrete breakouts lead to an inhomogeneous outflow composed of a series of hot, dense, and fast fronts of material that are trailed by gas which has expanded, cooled, and slowed down. This is reminiscent of what is observed in the M82 wind. 

\begin{figure}
\vspace*{-0.5cm}   
\includegraphics[width=0.525\textwidth]{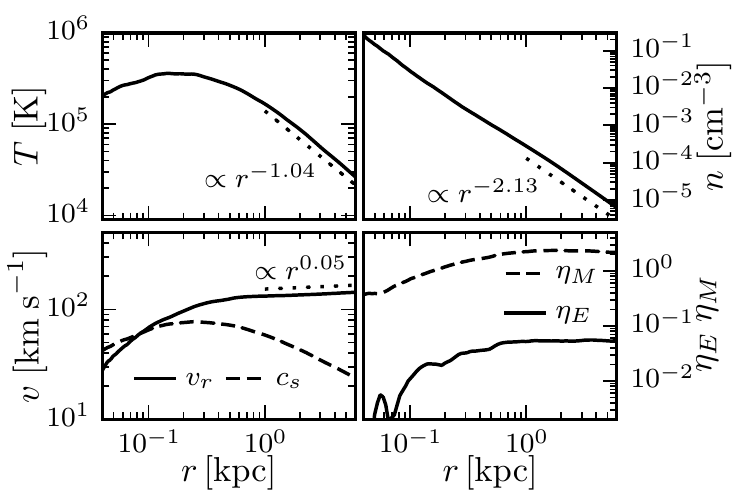}
\vspace*{-0.5cm}   
\caption{Time averaged radial profiles of $T$, $n$, $v_r$, $c_s$, $\etaE$, and $\etaM$ for the fiducial $\Sigmagas = 10~\Msunpc$ simulation. The averages are volume-weighted and computed in a biconical region with a $45^\circ$ half-opening angle. The best fit power-law slope between 1 and 6 kpc is shown for $T$, $n$, and $v_r$. The outflow properties are reasonably consistent with analytic models \citepalias{CC85}.}\label{fig:profiles}
\vspace*{-0.05cm}   
\end{figure}

In Fig. \ref{fig:profiles} we show the time averaged radial profiles of $T$, $n$, $v_r$, $c_s$, $\etaE$, and $\etaM$ for the fiducial $\Sigmagas {=} 10\,\Msunpc$ simulation. The averages are volume-weighted and computed in a biconical region centered on the disc rotation axis with a half opening angle of $45^\circ$. Several features are immediately apparent. The outflow is supersonic. The mass and energy outflow rates are roughly independent of radius beyond a certain point, indicating that we have a steady state outflow. The density of the wind material falls off rapidly, and the temperature decreases with radius slower than expected for just adiabatic expansion, which predicts $T\propto r^{-4/3}$. We have omitted the profiles from simulations with other parameters because they are all sufficiently similar and show the same trends.

Comparing the profiles in Fig. \ref{fig:profiles} to those from local Cartesian box simulations demonstrates the simulation geometry's effect on the wind structure. For example, Fig. 4 of \citetalias{Martizzi+16} shows that the sound speed and velocity of the wind are roughly independent of height beyond the scale height of the disc and the flow is always subsonic. Additionally, Fig. 9 and B1 of \citetalias{Martizzi+16} show that $\etaM$ decreases dramatically with distance from the disc whereas we find $\etaM$ to be roughly constant with radius. It is, therefore, the ratio of wind thermal to kinetic energy and the fraction of wind material that escapes that are primarily affected by the simulation geometry (important quantities for galactic winds!). 

Standard theoretical arguments for the structure of a galactic wind of a given $\etaE$, $\etaM$, and $\dot{\Sigma}_\star$ assume spherical symmetry and a uniform injection of energy and mass (e.g., \citetalias{CC85}, \citealt{Thompson+16}). Nevertheless, a comparison to the analytic work is instructive. A generic prediction of these models is that the wind will be supersonic beyond a sonic point that is approximately the radius of the star forming region (unless cooling is too strong). Our simulations agree with this prediction very well as can be seen in Fig. \ref{fig:profiles} where the sonic point is at ${\sim} R_d/3$. \citetalias{CC85} predict the asymptotic velocity to be $v_{\infty} \approx 10^3$ \kms $(\etaE/\etaM)^{1/2}$ and the temperature at the sonic point to be $T \approx 2\times10^7$K $\etaE/\etaM$, which agrees strikingly well with our simulations.  We indicate the best fit power law slope for the $T$, $n$, and $v_r$ profiles at large radii in Fig. \ref{fig:profiles} for comparison to observations and analytic models. Interestingly, $T$ falls off slower with distance than is expected for adiabatic expansion ($T\propto r^{-4/3}$), which could be due to cooling at small radii, additional heating beyond $R_d$ either by the rare (but effective) distant SNe, or by internal shocks. The biconical $n$ profile falls off roughly as $r^{-2}$ as expected for a freely expanding constant $\Mdot$ wind. When averaging over a spherical region the $n$ profile falls off much more quickly as $r^{-\alpha}$ with $\alpha{\sim}4$-$5$ for different simulations. The steeper fall off of the spherically averaged profile indicates there may be some fall back of wind material on the side of the biconical outflows creating a fountain flow (see Fig. \ref{fig:image}). This steeper slope is also consistent with the values inferred for local starburst galaxy M 82 \citep{Leroy+15b}.

\begin{figure}
\vspace*{-0.25cm}   
\includegraphics[width=0.48\textwidth]{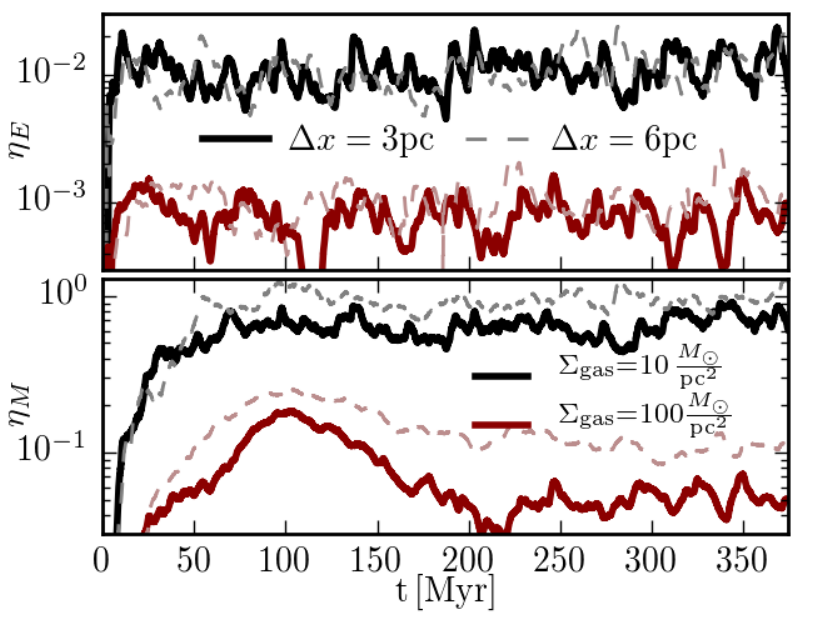}
\vspace*{-0.5cm}   
\caption{Time evolution of the energy outflow rate through a 4 kpc sphere normalized by SNe energy injection rate ($\etaE$), and the mass outflow rate through a 4 kpc sphere normalized to the star formation rate ($\etaM$) for the fiducial $\Sigmagas=10$ and 100 $\Msunpc$ simulations, shown in black and red, respectively. Identical simulations with half the resolution are shown with the thin dashed lines demonstrating the convergence of $\etaE$, but not $\etaM$.}\label{fig:outflow_evo_example}
\vspace*{-0.05cm}   
\end{figure}

In Fig. \ref{fig:outflow_evo_example} we show the time evolution of $\etaE$ and $\etaM$ for the fiducial $\Sigmagas {=} 10$ and $100~\Msunpc$ simulations at the highest and half the highest central spatial resolution. The energy and mass outflow rates of the wind are measured at 4 kpc. After the initial transient the outflow settles into a steady state (quite different from local Cartesian box simulations; e.g., \citetalias{Martizzi+16} Fig. 9 and B1). We performed extensive resolution testing on our simulations and found that with even half the resolution $\etaE$ is converged. However, as we increased the resolution $\etaM$ continued to decrease.
This is likely due to more mixing from unresolved shear layers with worse resolution, and more efficient venting at high resolution. The convergence of $\etaE$ does not depend sensitively on the degree to which we resolve the SNRs' $\rcool$. 
In our calculations, a resolved SNR has $\rcool {>} r_{\rm inj} {=} 2 \Delta x$. In the fiducial $\Sigmagas=10~\Msunpc$ simulation with $\Delta x = 3,6,$ and 12 pc 97.3, 73.5, and 31.8 per cent of the SNR are resolved, respectively, and yet $\etaE$ is the same. In the fiducial $\Sigmagas=100~\Msunpc$ simulation with $\Delta x = 3,6,$ and 12 pc 49.0, 12.1, and 1.4 per cent of the SNRs are resolved, respectively. In these higher surface density simulations $\etaE$ is roughly the same for the two higher resolution cases, but at $\Delta x = 12$ pc, $\etaE$ drops by a factor of ${\sim}2$. Overall, because $\etaM$ measured at large radii depends sensitively on the structure of the disc and the properties of the surrounding circumgalactic medium swept up by the wind \citep{Sarkar+15} we consider $\etaM$ of secondary importance relative to $\etaE$. Nevertheless, the lack of $\etaM$ convergence is something that should be considered whenever simulations similar to ours are compared to observations.

A standard physical picture of models such as \citetalias{CC85} is that winds are launched when the volume filling fraction of hot gas in the disc is large enough so that the number of SNe per cooling time and cooling volume---known as the porosity $Q_c$ \citep{McKeeOstriker77}---is greater than 1. \cite{Martizzi+15} found that 
$\rcool \approx  20.8~\mathrm{pc}~{n_H}^{-2/5}~\fcl^{2/7}$, 
and 
$t_{\rm cool} \approx 2.9\times10^{4}$ yrs ${n_H}^{-0.54}$.
Therefore, the porosity of the disc is
$Q_c = (4 \pi/3) \rcool^{3} t_{\rm cool} \dot{n} _{SN} 
= 6 \times 10^{-5} (n_H/100~\mathrm{cm}^{-3})^{-4/5}(100/f_\star) (10^6~\mathrm{yrs}/t_{\rm dyn})$.
In all of the simulations carried out here $Q_c \ll 1$. As can be seen in Fig. \ref{fig:image}, the SNe that contribute to the launching of the wind are the (rare) ones whose SNRs are able to breakout of the disc before radiating away their energy. Working under the assumption that SNRs that breakout satisfy $\rcool \gtrsim h$, where $h$ is the local scale height of the gaseous disc, we now provide a simple argument for the expected scaling of the wind properties with the disc and injection properties. 
There is a critical hydrogen number density
$
n_{\mathrm{crit}} = (h / 20.8 \mathrm{~pc})^{-5/2}
\fcl^{5/7} ~\mathrm{cm}^{-3}
$
that satisfies $\rcool = h$. Since, by design, the SN rate is proportional to the local density, the fraction of SNe that satisfy $n_H \leq n_{\mathrm{crit}}$ should be roughly equal to the ratio of $n_{\mathrm{crit}}$ to the midplane density $n_{\rm mid}~{\sim}~\Sigmagas / 2 h m_p$ (this need not be true for models with different SN seeding schemes or that account for additional physics).
Neglecting further radiative loses post breakout, the fraction of the injected energy that goes into the wind should be equal to the same ratio, which yields 
\be \label{eq:etaE} 
\etaE = \frac{\Edotw}{\Edot_{\rm SN}} \sim \frac{n_{\rm crit}}{n_{\rm mid}} \propto h^{-3/2} ~ \fcl^{5/7}  ~ \Sigmagas^{-1}. 
\ee
The total energy injection rate is
$ \Edot_{\rm SN} {=} 10^{51} \mathrm{~ergs~}  \pi R_d^2 \Sigmagas  / (t_\star \,100 \Msun),$
which follows from the definition of $t_\star$ and our model's assumption that there are $10^{51}$~ergs released per SN and there is one SN per $100 \Msun$ of stars formed. Combining this expression with equation (\ref{eq:etaE}) we find that 
$ 
\Edotw = \etaE \Edot_{\rm SN} \propto R_d^2 \, \fcl^{5/7} / (h^{3/2} \, t_\star).
$
One particularly interesting feature of this expression is the lack of any dependence on $\Sigmagas$. Finally, the ratio of the scale height to the disc radius is approximately equal the ratio of the velocity dispersion to the circular velocity ${h}/{R_d} \sim {\delta v}/{\vcirc}$, which is $\sim 0.1$ in our model. With this assumption in hand we end up with the following expected scalings for the wind energetics
\be
\etaE \propto R_d^{-3/2} \left(\frac{\delta v}{\vcirc}\right)^{-3/2} \fcl^{5/7} ~ \Sigmagas^{-1}.\label{eq:etaE_scalings}
\ee

\begin{figure}
\vspace*{-0.0cm}   
\includegraphics[width=0.475\textwidth]{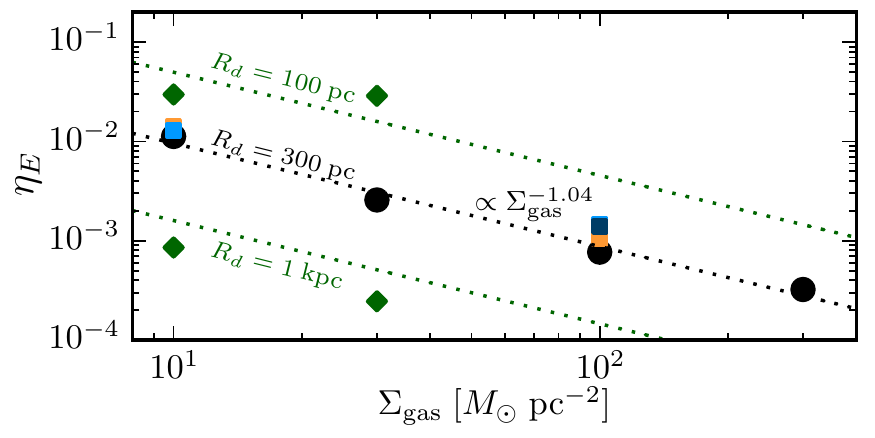}\vspace*{-0.05cm}   
\includegraphics[width=0.475\textwidth]{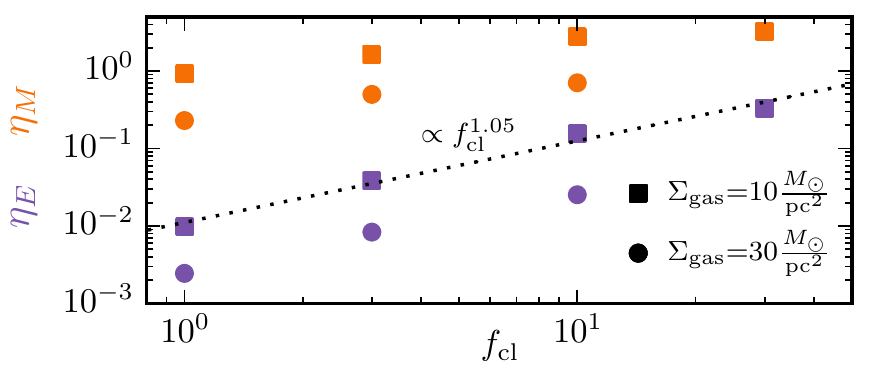}\vspace*{-0.05cm}   
\caption{(Top) The time averaged energy loading $\etaE$ versus $\Sigmagas$. The black circles correspond to simulations with $f_\star = 100$ and $\fcl$ = 1, colored squares correspond to different $f_\star$, which controls the star formation rate (orange is higher SFR, blue is lower), and the green diamonds show $\etaE$ for different disc sizes. The black dotted line shows the best fit to the fiducial models and the green dotted lines show the predicted $\etaE$ for different $R_d$ from our analytic model in which the wind properties are set by the SNe that breakout of the galactic disc (eq. \ref{eq:etaE_scalings}). (Bottom) The dependence of the time averaged $\etaE$ (purple) and $\etaM$ (orange) through a 4 kpc sphere on the degree of clustering $\fcl$ (\S \ref{Method}) for $\Sigmagas=10$ and $30\,\Msunpc$ shown with squares and circles, respectively. Clustering the SNe significantly boosts the wind energy and mass loss rates.}\label{fig:outflows}
\vspace*{-0.05cm}   
\end{figure}

The scalings above have been derived for $\etaE$ because the condition $\rcool \gtrsim h$ is explicitly a statement about energetics, but similar reasoning can be applied to mass-loading. Assuming that each SNR that breaks out contributes to the wind all of the mass it swept up prior breakout, $M_{\rm swept} = (4 \pi / 3) \rcool^3 \rho \approx 1200 ~ \Msun ~ n_{\rm crit}^{-1/5}$, yields $\etaM \approx (n_{\rm crit}/n_{\rm mid}) (M_{\rm swept} \, \dot{N}_{\rm SN} / \dot{M}_\star) = 12 \, \etaE \, n_{\rm crit}^{-1/5}$. Therefore, we expect $\etaM$ to scale similarly to $\etaE$ and to be one to two orders of magnitude larger than $\etaE$.

The top panel of Fig. \ref{fig:outflows} demonstrates that, for all other properties being equal, $\etaE$ is inversely dependent on $\Sigmagas$ exactly as predicted by equation (\ref{eq:etaE_scalings}). Increasing the star formation efficiency by decreasing $f_\star$ leads to an increase in $\Edotw$, but no appreciable change in $\etaE$ as expected. Likewise, equation (\ref{eq:etaE_scalings}) captures roughly the correct behavior for the scaling of $\etaE$ with $R_d$ seen in Fig. \ref{fig:outflows}, which demonstrates that more compact systems launch more powerful winds. 

The bottom panel of Fig. \ref{fig:outflows} shows how $\etaE$ and $\etaM$ scale with $\fcl$ for $R_d = 300$ pc  discs with $\Sigmagas=10$ and $30~\Msunpc$. In these simulations $\etaE$ increases roughly linearly with $\fcl$---somewhat more strongly than predicted in equation (\ref{eq:etaE}). 
Regardless of the exact scaling, the strong $\fcl$ dependence of $\etaE$ may be critical for understanding the launching of real galactic winds that are powerful enough to match observations and satisfy the requirements from cosmological simulations. This is because stars---massive stars in particular---are expected to form in clusters and it is likely that their SNRs will be nested or overlap rather than being spatially and temporally separated as in our $\fcl=1$ simulations. Although $\etaM$ also increases with $\fcl$, the measured scaling is less robust due to the lack of $\Mdotw$ convergence. 
Nevertheless, it is worth noting that for nearly all cases $\etaM$ surpasses 0.03---the value corresponding to all wind material coming from SNe ejecta.

\section{Discussion \& Conclusion} \label{Discussion}
Using idealized global galactic disc simulations we have quantified the properties of galactic winds driven solely by SNe for a range of disc and star formation properties. We have focused on small discs (${\sim} 0.1 {-} 1$ kpc in size) in order to ensure that the cooling radii of most of the SNRs in our simulations can be resolved. Our simulations roughly reproduce the supersonic wind structure expected from analytic models \citepalias[e.g.,][]{CC85}. Previous simulations that attempted to study galactic winds launched by SNe in a stratified medium often adopted local Cartesian domains (with periodic and outflow boundary conditions in the disc plane and perpendicular to it, respectively) and found subsonic outflows with outflow rates that depend on box height \citep[e.g.,][]{Martizzi+16,Girichidis+16,Kim+16}. The more physical global geometry we adopt allows the winds to adiabatically expand causing them accelerate to supersonic velocities, and the gravitational potential with a well-defined escape velocity leads to outflows with radially constant $\Mdotw$ and $\Edotw$ (Fig. \ref{fig:outflow_evo_example}). Other numerical models for studying galactic winds inject a fixed $\Edot$ uniformly in a given volume \citep{StricklandHeckman09,Sarkar+15}. Our calculations compliment these by addressing the key question of how discrete SNe collectively drive a wind.

In analytic galactic wind models the mass ($\etaM$) and energy ($\etaE$) loading of the wind are free parameters. Our simulations determine these wind properties as a function of the underlying disc structure (e.g., $\Sigmagas$ and $R_d$) and the SN seeding model (e.g., degree of clustering $\fcl$).
In our simulations the winds are driven by SNe that go off in low density regions where the cooling radius $\rcool$ is larger than the local scale height $h$; this enables SNRs to drive the wind without radiative losses sapping their energy. In general only a small fraction of the SNe satisfy this constraint because at the disc midplane $\rcool \ll h$. We present a simple model based on this concept (see eq. \ref{eq:etaE_scalings}) that predicts, among other things, that $\Edotw$ should be independent of $\Sigmagas$, increase with the degree of SNe clustering $\fcl$ and the star formation efficiency $f_\star$, and decrease with increasing disc size. This simple model successfully explains many of the trends we find in our simulations (Fig. \ref{fig:outflows}). Although this analytic model and the numerical scaling of wind efficiency with disc and SNe parameters are likely to be somewhat modified with different SNe seeding schemes, we expect that the general trends found here are likely to be more robust---in particular the simple criterion that $\rcool \gtrsim h$ for the SNe that drive the wind.

The mass and energy loading of the galactic winds driven by SNe we find are likely lower than suggested by observations. This may be due to the fact that our SNe are set off preferentially at density peaks and that the ISM is relatively homogeneous. A more realistic (or a spatially random) SNe seeding scheme that separates the SN locations from density peaks and/or clustering the SNe would increase the efficiency of the outflows \citep[e.g.,][]{Sharma+14,Girichidis+16,Kim+16,Gentry+17}. 
Indeed, in our calculations, we implemented a simple model of clustering in which each SNR's energy is increased by a factor of $f_{\rm cl}$, and the SN rate decreased by the same factor, leaving the total energy input rate the same.  The resulting galactic wind energy loss rate increases roughly linearly with $f_{\rm cl}$ (Fig. \ref{fig:outflows}). The wind power might well be further increased if additional physics were included such as molecular line cooling \citep{Li+16} and stronger ISM turbulence possibly enhanced by self-gravity and/or galactic inflows \citep{Sur+16}. These would result in larger density inhomogeneities, causing more of the ISM volume to be filled with low density gas, and thereby allowing more SNe to go off in low density regions and breakout of the disc. 
\vspace*{-0.2cm}   

\section*{Acknowledgements}
We thank Prateek Sharma for useful conversations during the development of this work. This work was supported in part by NASA ATP grant 12-APT12-0183 and a Simons Investigator award from the Simons Foundation to EQ. DF was supported by the NSF GRFP under Grant \# DGE 1106400.  CAFG was supported by NSF through grants AST-1412836 and AST-1517491, and by NASA through grant NNX15AB22G. DM was supported by the Swiss National Science Foundation as an Advanced Postdoc Mobility Fellow until Nov. 2016; grant number P300P2\_161062.

This research used the Savio computational cluster resource provided by the Berkeley Research Computing program at UC, Berkeley (supported by the UC Berkeley Chancellor, Vice Chancellor of Research, and Office of the CIO). In addition, this work used the Extreme Science and Engineering Discovery Environment (XSEDE), which is supported by National Science Foundation grant number ACI-1053575, via grant number TG-AST160020.

\bibliographystyle{mnras}
\bibliography{Supernovae}

\label{lastpage}

\end{document}